\pdfoutput=1

\newcommand{\average}[1]{\mbox{$\langle #1 \rangle$}}
\newcommand{\ket}[1]{\mbox{$ | #1 \rangle $}}
\newcommand{\bra}[1]{\mbox{$ \langle #1 | $}}

\newcommand{\Id}{\mathds{1}}
\newcommand{\beq}{\begin{eqnarray}}
\newcommand{\eeq}{\end{eqnarray}}

\newcommand{\tr}{\mbox{Tr}\,}

\newcommand{\ps}{\ket{\psi}}
\newcommand{\Ah}{\hat{A}}
\newcommand{\Bh}{\hat{B}}

\documentclass[pra,onecolumn,superscriptaddress]{revtex4}

\usepackage[T1]{fontenc}
\usepackage[utf8]{inputenc}
\usepackage{amsmath}
\usepackage{amsfonts}
\usepackage{amsthm}
\usepackage{amssymb}
\usepackage{subeqnarray}
\usepackage{dsfont} 
\usepackage{setspace}
\usepackage{graphics}
\usepackage{url}
\usepackage{hyperref}
\usepackage{ifthen}
\usepackage{color}
\usepackage{graphicx}
\usepackage{enumitem}
\usepackage{float}
\usepackage{natbib} 
\usepackage{multirow}
\usepackage{hhline}
\usepackage[makeroom]{cancel}
\usepackage[scr=boondoxo,scrscaled=1.05]{mathalfa}

\newtheorem{lem}{Lemma}
\newtheorem{theorem}{Theorem}
\newtheorem{defn}{Definition}

\begin{document}

\title{All Pure Bipartite Entangled States can be Self-Tested}

\author{Andrea Coladangelo}
\affiliation{Department of Computing and Mathematical Sciences, California Institute of Technology,
1200 E California Blvd, Pasadena, CA 91125, United States}

\author{Koon Tong Goh}
\affiliation{Centre for Quantum Technologies, National University of Singapore, 3 Science Drive 2, Singapore 117543}

\author{Valerio Scarani}
\affiliation{Centre for Quantum Technologies, National University of Singapore, 3 Science Drive 2, Singapore 117543}
\affiliation{Department of Physics, National University of Singapore, 2 Science Drive 3, Singapore 117542}

\begin{abstract}
Device-independent self-testing allows to uniquely characterize the quantum state shared by untrusted parties (up to local isometries) by simply inspecting their correlations, and requiring only minimal assumptions, namely a no-signaling constraint on the untrusted parties and the validity of quantum mechanics. The device-independent approach exploits the fact that certain non-local correlations can be uniquely achieved by measurements on a particular quantum state. We can think of these correlations as a ``classical fingerprint'' of the self-tested quantum state.
In this work, we answer affirmatively the outstanding open question of whether all pure bipartite entangled states can be self-tested, by providing explicit self-testing correlations for each. 
\end{abstract}

\maketitle
%
\section{Introduction}
Device-independent self-testing enables a completely classical verifier to characterize the joint quantum state shared by two potentially untrusted parties (the provers), up to local isometries, by simple inspection of the observed correlations. This approach requires minimal assumptions, namely a no-signaling constraint on the provers, and the validity of quantum mechanics. Thus, in a device-independent scenario \cite{Acin2007}, one can obtain guarantees on the functionality of a device without making assumptions on its inner-workings.

Self-testing is made possible by the existence of non-local correlations in quantum theory. While all correlations produced by classical provers are necessarily local, on the other hand, it is possible to produce non-local correlations by measuring a joint quantum state that is entangled \cite{Bell1964}. It is well-known that certain entangled quantum states can be self-tested, meaning that a classical verifier can certify that such a state is shared by the two provers by observing the maximal violation of some Bell inequality, the ideal winning probability in some non-local game played by the provers, or simply by observing correlations that can be uniquely obtained by measurements on that state. The most celebrated example of a state that can be self-tested is the maximally entangled pair of qubits (the singlet). This is self-tested, for instance, by the maximal violation of the well-known Clauser-Horne-Shimony-Holt (CHSH) inequality \cite{SW87, Popescu92}. 

The term ``self-testing'' in the context of Bell experiments was coined by Mayers and Yao \cite{Mayers2004}, who introduced a simple criterion to self-test the singlet state for the bipartite Bell scenario with three dichotomic measurements on each side. Since then, self-testing of the two-qubit singlet has been made robust \cite{MYS12}, then extended to sequential \cite{Reichardt2013} and parallel certification of many copies \cite{Wu2016,Coladangelo2016,Coudron2016,Mckague2016,Natarajan2016,CRSV2016}; and the complete set of criteria that self-test that state with two dichotomic measurements has been provided \cite{Wang2016}. Moreover, a variety of other quantum states have been proved to be self-testable: all partially entangled pure two-qubit states \cite{Yang2013,Bamps2015}, the maximally entangled pair of qutrits \cite{Salavrakos2016}, the partially entangled pair of qutrits that violates maximally the CGLMP$_3$ inequality \cite{Acin2002,Yang2014}, and a small class of higher dimensional partially entangled pairs of qudits, through results in parallel self-testing \cite{Coladangelo2016}. For the multi-partite case, it is known that the three-qubit W state \cite{Wu2014,Pal2014} and graph states \cite{McKague2011,Pal2014} can be self-tested. Hence, it is clear that self-testing is not an exclusive characteristic of maximally entangled states nor qubit states. However, little is known about self-testing of higher-dimensional entangled states (i.e. pairs of entangled qudits for $d >2$).

In this paper, we consider the outstanding open question of whether all bipartite pure entangled quantum states can be self-tested. Building on the framework of Yang and Navascu\'es \cite{Yang2013}, we answer this question affirmatively with an explicit construction of a family of self-testing correlations, with $3$ and $4$ measurement settings for the two provers respectively, and $d$ outcomes per party (where $d$ is the local dimension). We argue, additionally, that our correlations self-test not only the state, but also certain ideal measurements.

On top of answering a fundamental question of quantum information science, our result has potential applications in cryptography. Several known protocols are in fact based on self-testing, usually on the rigidity of the CHSH game \cite{Reichardt2013, Coudron2013,MillerShi2016}, and our work adds the flexibility of choosing the self-tested state, and certain corresponding $d$-outcome measurements, in the bipartite scenario. One potential application is in device-independent randomness expansion, the first device-independent Quantum Random Number Generation scheme to be proposed and the only to have been experimentally implemented \cite{Pironio2010}. There, guaranteed private randomness is generated from an initial random seed. Based on our self-testing result, one could hope to generate up to $O(\log d)$ bits of private randomness, with $d$ limited only by the experimental state-of-the-art, using a small random seed (two random trits per run). An analysis of the robustness of our result is required in order to assess its applicability, and we leave this exploration for future work. 

The paper is organized as follows. Section \ref{preliminaries} contains preliminary notions and, in particular, a criterion giving sufficient conditions for self-testing an entangled pair of qudits. Section \ref{section_correlations} contains the main result, a description of the self-testing correlations and the ideal measurements achieving them. Section \ref{proof} contains a proof of the main result. Sections \ref{discussion} and \ref{conclusion} contain a brief discussion and conclusion.

\section{Preliminaries}
\label{preliminaries}
\subsection{General framework}
\begin{figure}[H]
  \centering
    \includegraphics[width=0.55\textwidth]{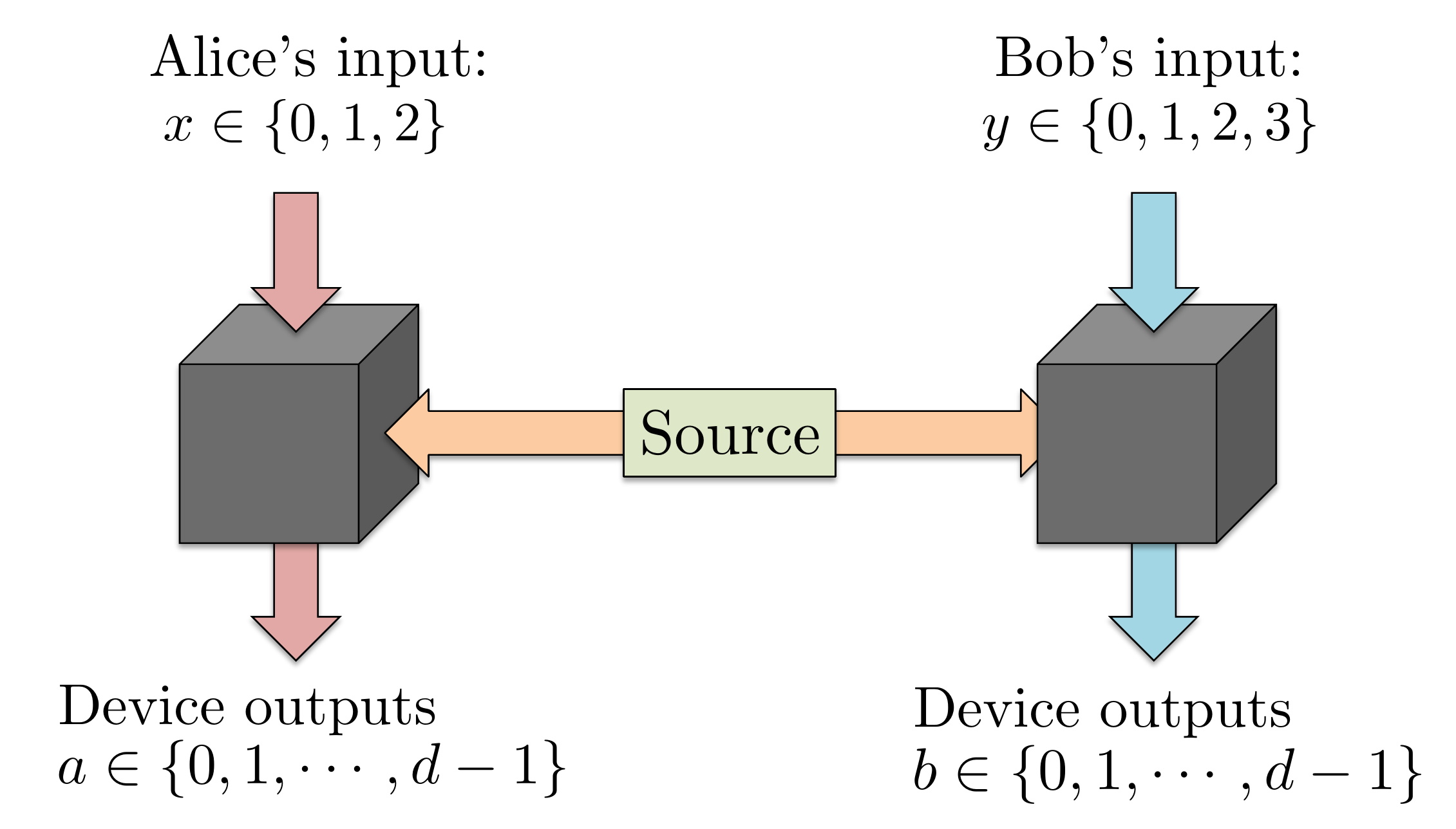}
      \caption{A [$\{3,d\},\{4,d\}$] Bell scenario}
      \label{fig:bell}
\end{figure}

In a bipartite Bell scenario, the two provers Alice and Bob receive inputs $x \in \mathcal{X}$ and $y \in \mathcal{Y}$ respectively, corresponding to their choice of measurement settings, and their devices return outcomes $a \in \mathcal{A}$ and $b \in \mathcal{B}$.\\
We refer to the collection of conditional probability distributions $$\{P(a,b|x,y): (a,b) \in \mathcal{A} \times \mathcal{B}\}_{(x,y) \in \mathcal{X}\times \mathcal{Y}}$$
as a \textit{correlation} (where we sometimes interchangeably utilize the plural of the word when the context leaves no ambiguity). 
In the particular scenario that we consider for our self-testing correlations, Alice has three possible measurement settings and Bob has four, while the devices have $d$ possible outcomes. So the inputs are $x\in\{0,1,2\}$ and $y\in\{0,1,2,3\}$, and the outputs are $a,b \in \{0,1,2,\cdots,d-1\}$. We refer to this as a $[\{3,d\},\{4,d\}]$ Bell scenario (see FIG.~\ref{fig:bell}), and we denote by $\mathcal{C}_q^{3,4,d,d}$ the set of quantum correlations that two parties can generate in such a Bell scenario. 

\noindent A convenient way to represent correlations is through correlation tables: for the $[\{3,d\},\{4,d\}]$, we can arrange the $P(a,b|x,y)$ in twelve $d\times d$ correlation tables, one for each pair of measurement settings, denoted by $T_{x,y}$.
\begin{equation}
\label{Txy}
T_{x,y}:=
\begin{tabular}{|c||c|c|c|c|}
    \hline
    $a \backslash b$ & 0 & 1 & $\cdots$ & $d-1$ \\ \hline \hline
    0 & $P(0,0|x,y)$ & $P(0,1|x,y)$ & $\cdots$ & $P(0,d-1|x,y)$ \\ \hline
    1 & $P(1,0|x,y)$ & $P(1,1|x,y)$ & $\cdots$ & $P(1,d-1|x,y)$ \\ \hline
    $\vdots$ & $\vdots$ & $\vdots$ & $\ddots$ & $\vdots$ \\ \hline
    $d-1$ & $P(d-1,0|x,y)$ & $P(d-1,1|x,y)$ & $\cdots$ & $P(d-1,d-1|x,y)$ \\
     \hline
\end{tabular}
\end{equation}

\begin{defn}(Self-testing)
We say that a correlation \textit{self-tests} a state $\ket{\Psi}$ and measurement operators $\{A_x\}$, $\{B_y\}$ on Alice and Bob's side, if it can be reproduced uniquely when the devices of Alice and Bob make measurements $\{A_x\}$, $\{B_y\}$ on the joint state $\ket{\Psi}$, up to a local isometry.
\end{defn}

In the device-independent approach, the dimensionality of the measured system is not bounded a priori. Hence, the measurements made on the system can be assumed to be projective, thanks to Naimark's theorem, with $\Pi^{A_x}_a$ the projection corresponding to Alice obtaining outcome $a$ on measurement setting $x$, and likewise for $\Pi^{B_y}_b$ on Bob's side. We don't have to assume that the unknown joint state shared by Alice and Bob is pure, but rather we take it as such for ease of exposition, and we denote it by $\ket{\psi}$. It is then easy to see that our proof holds through in the same way for a general joint state $\rho$. Notice, finally, that any correlations produced by a bipartite mixed state can be reproduced by a bipartite pure state of the same dimension \cite{Sikora2015}, which implies that bipartite mixed states cannot be self-tested. Hence, in the bipartite scenario, the best one can hope for is to self-test every pure state, and this is what we obtain in the present work. No further characterization of either the state or the measurements is required, and estimating the $P(a,b|x,y)$ is all that has to be done in the lab. 

Our objective is to self-test all bipartite quantum states, and, using the Schmidt decomposition, this reduces to self-testing an arbitrary bipartite quantum state of the form
\begin{equation}
\ket{\psi_{\text{target}}}:=\sum_{i=0}^{d-1} c_i \ket{ii}
\end{equation}
where $0 < c_i < 1$ for all $i$ and $\sum_{i=0}^{d-1}c_i^2 = 1$. \\ 

\subsection{Tilted CHSH inequality}
We briefly introduce the tilted CHSH inequality \cite{Acin2012}, which will be a building block for our self-testing correlations. Let $A_0, A_1, B_0, B_1$ be $\pm 1$ random variables. For a random variable $X$, let $\left<X\right>$ denote its expectation. The term \textit{tilted CHSH inequality} refers to a one-parameter family of inequalities, which generalises the CHSH inequality:
\begin{equation}
    \left<\alpha A_0 + A_0B_0 + A_0B_1 +A_1B_0 - A_1B_1 \right > \leq 2+\alpha
    \label{tiltedchsh}
\end{equation}
which holds when the random variables are local. The maximal quantum violation is $\sqrt{8+2\alpha^2}$, and is attained when $A_0, A_1, B_0, B_1$ are the binary observables $A_0 = \sigma_Z$, $A_1 = \sigma_X$, $B_0 = \cos \mu \sigma_Z + \sin \mu \sigma_X$ and $B_1 = \cos \mu \sigma_Z + \sin \mu \sigma_X$, and the underlying joint state is $\ket{\psi} = \cos \theta \ket{00} + \sin \theta \ket{11}$, where $\sin 2\theta = \sqrt{\frac{4-\alpha^2}{4+\alpha^2}}$, $\mu = \arctan (\sin 2\theta)$, and $\sigma_Z$, $\sigma_X$ are usual Pauli matrices. The converse also holds, in the sense that maximal violation self-tests this state and these measurements \cite{Bamps2015}. 

The following Lemma from \cite{Bamps2015} will be useful later on. In what follows, a subscript indicates the subsystem that an operator acts on ($A$ for Alice and $B$ for Bob). When it is clear from the context, we omit writing trivial identities on other subsystems: for example, we may write $A_0$ in place of $A_0 \otimes \mathds{1}$.
\begin{lem}
\label{Bamps lemma}
Let $\ket{\psi} \in \mathcal{H}_A \otimes \mathcal{B}$. Let $A_0, A_1$ and $B_0, B_1$ be binary observables, respectively on $\mathcal{H}_A$ and $\mathcal{H}_B$, with $\pm1$ eigenvalues. Suppose that
\begin{equation}
\bra{\psi}\alpha A_0 + A_0B_0 + A_0B_1 +A_1B_0 - A_1B_1 \ket{\psi} = \sqrt{8+\alpha^2}
\end{equation}
Let $\theta, \mu \in (0,\frac{\pi}{2})$ be such that $\sin 2\theta = \sqrt{\frac{4-\alpha^2}{4+\alpha^2}}$ and $\mu = \arctan \sin 2\theta$. Then, let $Z_A = A_0$, $X_A = A_1$. Let $Z^*_B$ and $X^*_B$ be respectively $\frac{B_0 + B_1}{2\cos \mu}$ and $\frac{B_0 - B_1}{2\sin \mu}$, but with all zero eigenvalues replaced by one. Then, define $Z_B = Z^*_B|Z^*_B|^{-1}$ and $X_B = X^*_B|X^*_B|^{-1}$. \\
Then, we have
\begin{align}
Z_A\ps &= Z_B \ps \\
\cos\theta X_A (\mathds{1}-Z_A) \ps &=  \sin\theta X_B (\mathds{1}+Z_A) \ps 
\end{align}
\end{lem}

\subsection{Sufficient conditions for self-testing an entangled pair of qudits}

We state a (slightly more general) version of a Lemma from Yang and Navascu\'es \cite{Yang2013}, which gives a sufficient criterion for self-testing a general pure bipartite entangled state. 
\begin{lem}
\label{YNcriterion}
Let $\ket{\psi_{\text{target}}} = \sum_{i=0}^{d-1} c_i \ket{ii}$, where $0< c_i < 1$ for all $i$ and $\sum_{i=0}^{d-1}c_i^2 = 1$. Suppose there exist unitary operators $X^{(k)}_{A}, X^{(k)}_{B}$ and projections $\{P^{(k)}_{A}\}_{k=0,..,d-1}$ and $\{P^{(k)}_{B}\}_{k=0,..,d-1}$ of which $\{P^{(k)}_{A}\}_{k=0,..,d-1}$ is a complete orthogonal set, while $\{P^{(k)}_{B}\}_{k=0,..,d-1}$ need not be, and they satisfy the following conditions:
\begin{align}  
P^{(k)}_{A}\ket{\psi}&=P^{(k)}_{B}\ket{\psi} \;\; \forall k,  \label{c2} \\
X^{(k)}_{A}X^{(k)}_{B}P^{(k)}_{B}\ket{\psi}&=\frac{c_k}{c_0} P^{(0)}_{A}\ket{\psi}\,\,\forall k \label{c3}
\end{align}
Then there exists a local isometry $\Phi$ such that $\Phi(\ket{\psi})=\ket{\text{extra}}\otimes\ket{\psi_{\text{target}}}$, for some auxiliary state $\ket{\text{extra}}$.
\end{lem}
\begin{figure}[H]
  \centering
    \includegraphics[width=0.55\textwidth]{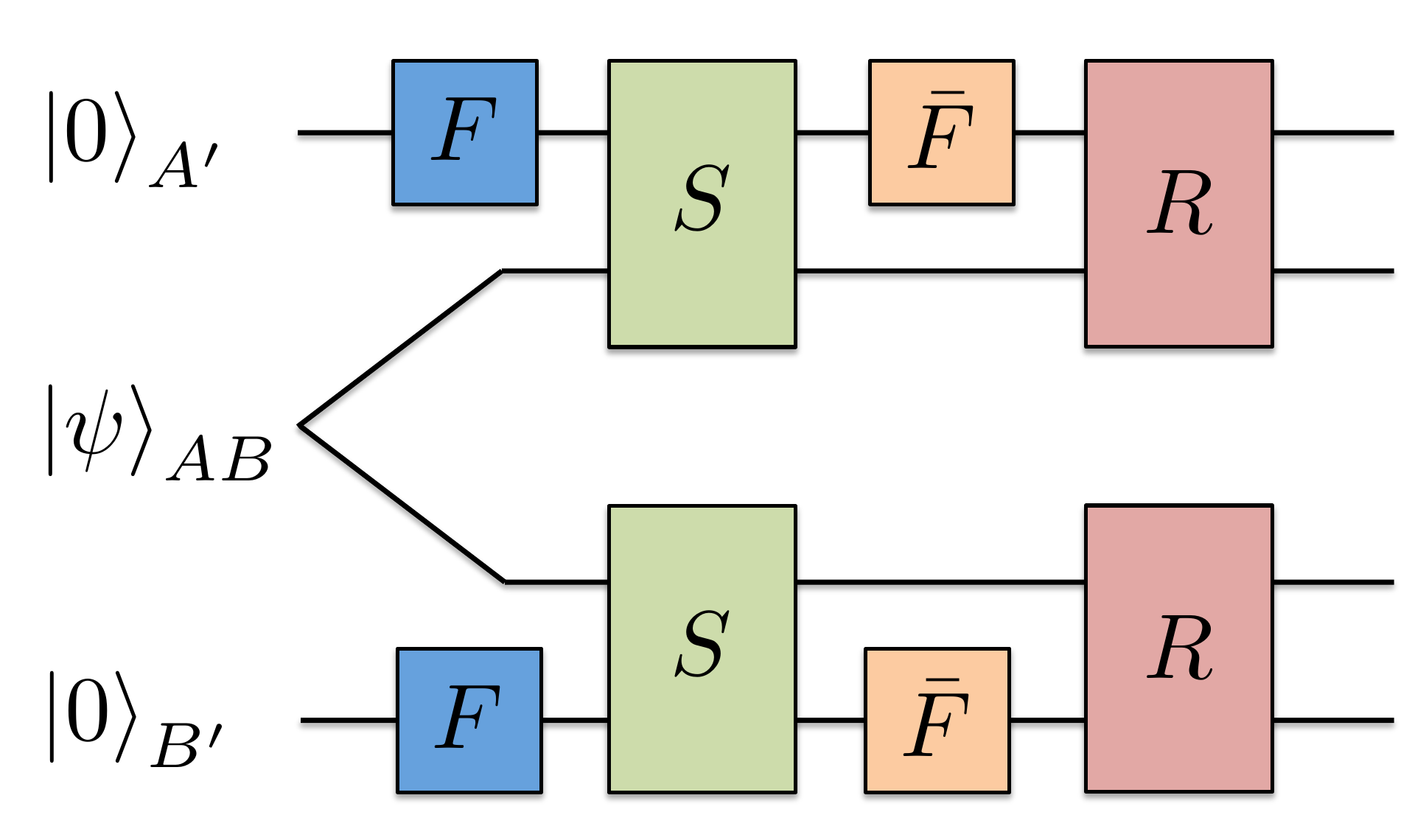}
      \caption{Diagram of the isometry $\Phi(\ket{\psi})$}
      \label{fig:iso}
\end{figure}

The complete proof of this is given in Appendix \ref{YNproof}. The Lemma also holds when $\ps$ is replaced by a general mixed state $\rho$, and equalities between vectors are naturally replaced by equalities between density matrices, as is clear from the proof in Appendix \ref{YNproof}. Here we just describe how the local isometry $\Phi$ is constructed (Fig. \ref{fig:iso}). The local isometry adds two ancilla qudits in the zero state, and is a generalization of the SWAP isometry that is typically used in the qubit case. More precisely, 


\begin{equation}
\Phi(\ket{\psi}) = (R_{AA'}\otimes R_{BB'})(\bar{F}_{A'}\otimes\bar{F}_{B'})(S_{AA'}\otimes S_{BB'})(F_{A'}\otimes F_{B'})\ket{\psi}_{AB}\ket{0}_{A'}\ket{0}_{B'} \label{iso}
\end{equation}
where $F$ is the quantum Fourier transform, $\bar{F}$ is the inverse quantum Fourier transform, $R_{AA'/BB'}$ is defined as $R_{AA'/BB'}\ket{\psi}_{AB}\ket{k}_{A'/B'}= X^{(k)}_{A/B}\ket{\psi}_{AB}\ket{k}_{A'/B'}$ and $S_{AA'/BB'}$ is defined as $S_{AA'/BB'}\ket{\psi}_{AB}\ket{k}_{A'/B'}= Z^{k}_{A/B}\ket{\psi}_{AB}\ket{k}_{A'/B'}$. Yang and Navascu\'es \cite{Yang2013} did not provide, or prove the existence of, correlations from which one can construct operators satisfying the conditions of Lemma \ref{YNcriterion}, and this is our main contribution.

\section{Self-testing correlations}
\label{section_correlations}
We first state our main theorem, then give a sketch idea of the self-testing correlations in subsection \ref{idea correlations}, which we describe precisely in subsection \ref{correlations}. In subsection \ref{ideal measurements}, we describe ideal measurements achieving the correlations. 

\subsection{Our main result}
\begin{theorem}
\label{main theorem}
For every bipartite entangled state of qudits $\ket{\psi_{\text{target}}}$, there exist correlations in $\mathcal{C}_{q}^{3,4,d,d}\,$ such that, when reproduced by Alice and Bob through local measurements on a joint state $\rho$, imply the existence of a local isometry $\Phi$ such that $\Phi(\rho) = \rho_{\text{extra}}\otimes\ket{\psi_{\text{target}}}\bra{\psi_{\text{target}}}$, where $\rho_{\text{extra}}$ is some auxiliary state. Moreover, under the isometry $\Phi$, the local measurements on $\rho$ are equivalent to measurements that act trivially on $\rho_{\text{extra}}$ and as the ideal measurements on $\ket{\psi_{\text{target}}}$ (described in subsection \ref{ideal measurements}).
\end{theorem}

The correlations that make Theorem \ref{main theorem} true are the ones we are about to describe.

\subsection{The idea behind our correlations}
\label{idea correlations}
\begin{figure}[H]
  \centering
    \includegraphics[width=0.55\textwidth]{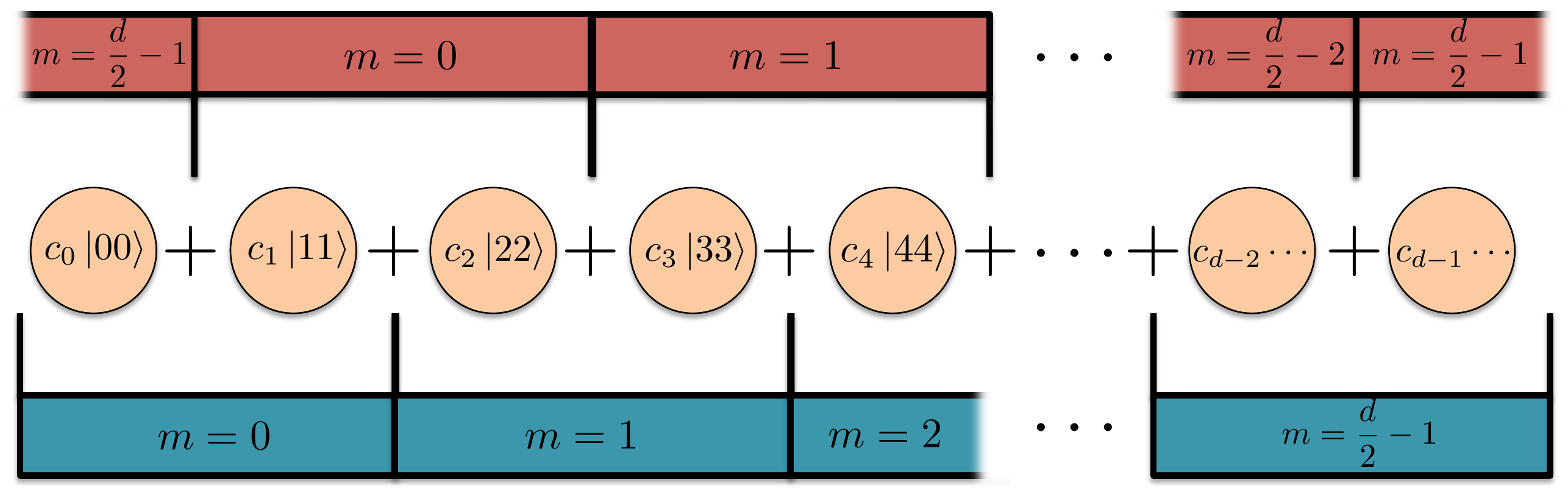}
      \caption{In blue, the block-diagonal correlations for measurement settings $x,y \in \{0,1\}$ ``certify'' the ``even-odd'' pairs, while, in red, the block-diagonal correlations for measurement settings $x \in \{0,2\}, y \in \{2,3\}$ certify the ``odd-even'' pairs.}
      \label{fig:intuition}
\end{figure}

Here we give a sketch of the structure of our self-testing correlations that we will describe in full detail in the following subsection, and an intuition of why they work. For clarity, in this paragraph we assume $d$ to be even, but the proof will apply to odd $d$ as well.\\
Recall that we wish to self-test the state $\ket{\psi_{\text{target}}} = \sum_{i=0}^{d-1} c_i \ket{ii}$, where $0 < c_i < 1$ for all $i$ and $\sum_{i=0}^{d-1}c_i^2 = 1$. The approach, inspired by \cite{Yang2013}, is to use $d$-outcome measurements on Alice and Bob's side such that, for some measurement settings, the correlation tables $T_{x,y}$, as defined in \eqref{Txy}, are block-diagonal with $2\times2$ blocks. More precisely, for measurement settings $x,y \in \{0,1\}$, the $2\times2$ blocks will correspond to outcomes $a,b$ respectively in $\{0,1\}$, in $\{2,3\}$,.., in $\{d-2,d-1\}$; and the idea is that the $m$th $2\times2$ block self-tests the portion $c_{2m}\ket{2m\,\,2m } + c_{2m+1}\ket{2m+1 \,\,\,2m+1}$ of the target state. For measurement settings $x \in \{0,2\}, y \in \{2,3\}$, instead, the $2\times2$ blocks will correspond to outcomes $a,b$ respectively in  $\{1,2\}$, in $\{3,4\}$,.., in $\{d-1,0\}$, again the idea being that the $m$th block ``self-tests" the portion $c_{2m+1}\ket{2m+1\,\,2m+1} + c_{2m+2}\ket{2m+2 \,\,\,2m+2}$.\\ See Fig. \ref{fig:intuition} for an illustration of the concept.

As one can expect, the $2\times 2$ blocks in our block-diagonal correlations will naturally correspond to ideal tilted CHSH correlations for appropriately chosen angles.

As we shall clarify later, this particular choice for the $2\times2$ blocks is not essential: although no other criterion for self-testing arbitrary partially entangled qubits is currently known, any other self-testing correlations from which we can deduce the existence of operators satisfying \eqref{f1} and \eqref{f2} could in principle be used in our proof.

\subsection{The correlations}
\label{correlations}
In order to self-test the target state $\ket{\psi_{\text{target}}} = \sum_{i=0}^{d-1} c_i \ket{ii}$, where $0 < c_i < 1$, we will not need to specify the entire set of twelve correlation tables $T_{x,y}$, but it will be sufficient to specify just the tables corresponding to measurement settings  $x,y \in \{0,1\}$, and those for settings $x \in \{0,2\}, y \in \{2,3\}$. We will show that any correlation satisfying these constraints self-tests $\ket{\psi_{\text{target}}}$.\\ Moreover, the constraints that we place on the correlations are sufficient to determine a single correlation (i.e. a single point in a quantum correlation set), and this is because such constraints self-test not only the state, but also ideal measurements (see subsections \ref{ideal measurements} and Appendix \ref{self-testing measurements}). We will often refer to the correlation achieved by the ideal measurements as the \textit{ideal correlation}. \\ 
The reader may find the description of the ideal measurements achieving these constraints, from subsection \ref{ideal measurements}, helpful in visualizing the ideal correlation. 

Building on an idea of Yang and Navascu\'es \cite{Yang2013}, the constraints that we impose on the correlations are:

\begin{enumerate}
\item[(i)] For $x,y \in \{0,1\}$, the correlation tables are block diagonal with $2\times 2$ blocks. The tables for measurement settings $x,y \in \{0,1\}$ are given in Tables \ref{tab:txy1} and \ref{tab:txy2} for even and odd $d$ respectively. The $2 \times 2$ blocks $C_{x,y,m}$ are given by $(c_{2m}^2+c_{2m+1}^2)\cdot C^{\text{ideal}}_{x,y,\theta_m}$ where the $C^{\text{ideal}}_{x,y,\theta_m}$ are the $2 \times 2$ correlation tables which correspond to the maximal violation of the tilted-CHSH inequality which self-tests the state $\cos{(\theta_m)}\ket{00}+\sin{(\theta_m)}\ket{11}$, where $\theta_m :=\arctan \big(\frac{c_{2m+1}}{c_{2m}}\big) \in (0,\frac{\pi}{2})$. They are given precisely in Tables~\ref{tab:corr1}-\ref{tab:corr4}, with $\mu_m := \arctan{(\sin{(2\theta_m)})}$. 
\begin{table}[H]
\caption{$T_{x,y}$ for $x,y\in \{0,1\}$ for even values of $d \geq 2$}
\label{tab:txy1}
\begin{center}
\begin{tabular}{| c || c | c | c | c | c | c | c |}
	\hline
	$a \backslash b$ & 0 & 1 & 2 & 3 & $\cdots$ & $d-2$ & $d-1$ \\ \hline \hline
	0 & \multicolumn{2}{| c |}{\multirow{2}{*}{$C_{x,y,m=0}$}} & 0 & 0 & $\cdots$ & 0 & 0 \\ 
	\hhline{*{1}{|-}*{2}{|~}*{5}{|-}}
	1 & \multicolumn{2}{| c |}{} & 0 & 0 & $\cdots$ & 0 & 0 \\ \hline
	2 & 0 & 0 & \multicolumn{2}{| c |}{\multirow{2}{*}{$C_{x,y,m=1}$}} & $\cdots$ & 0 & 0 \\
	\hhline{*{3}{|-}*{2}{|~}*{3}{|-}}
	3 & 0 & 0 & \multicolumn{2}{| c |}{} & $\cdots$ & 0 & 0 \\ \hline
	$\vdots$ & $\vdots$ & $\vdots$ & $\vdots$ & $\vdots$ & $\ddots$ & $\vdots$ & $\vdots$  \\ \hline
	$d-2$ & 0 & 0 & 0 & 0 & $\cdots$ &  \multicolumn{2}{| c |}{\multirow{2}{*}{$C_{x,y,m=\frac{d}{2}-1}$}} \\
	\hhline{*{6}{|-}*{2}{|~}}
	$d-1$ & 0 & 0 & 0 & 0 & $\cdots$ & \multicolumn{2}{| c |}{} \\ 
	\hline
\end{tabular}
\end{center}
\end{table}
\begin{table}[H]
\caption{$T_{x,y}$ for $x,y\in \{0,1\}$ for odd values of $d \geq 3$}
\label{tab:txy2}
\begin{center}
\begin{tabular}{| c || c | c | c | c | c | c | c | c |}
	\hline
	$a \backslash b$ & 0 & 1 & 2 & 3 & $\cdots$ & $d-3$ & $d-2$ & $d-1$ \\ \hline \hline
	0 & \multicolumn{2}{| c |}{\multirow{2}{*}{$C_{x,y,m=0}$}} & 0 & 0 & $\cdots$ & 0 & 0 & 0\\ 
	\hhline{*{1}{|-}*{2}{|~}*{6}{|-}}
	1 & \multicolumn{2}{| c |}{} & 0 & 0 & $\cdots$ & 0 & 0 & 0\\ \hline
	2 & 0 & 0 & \multicolumn{2}{| c |}{\multirow{2}{*}{$C_{x,y,m=1}$}} & $\cdots$ & 0 & 0 & 0\\
	\hhline{*{3}{|-}*{2}{|~}*{4}{|-}}
	3 & 0 & 0 & \multicolumn{2}{| c |}{} & $\cdots$ & 0 & 0 & 0\\ \hline
	$\vdots$ & $\vdots$ & $\vdots$ & $\vdots$ & $\vdots$ & $\ddots$ & $\vdots$ & $\vdots$  & 0\\ \hline
	$d-3$ & 0 & 0 & 0 & 0 & $\cdots$ &  \multicolumn{2}{| c |}{\multirow{2}{*}{$C_{x,y,m=\frac{d-3}{2}}$}} & 0\\
	\hhline{*{6}{|-}*{2}{|~}*{1}{|-}}
	$d-2$ & 0 & 0 & 0 & 0 & $\cdots$ & \multicolumn{2}{| c |}{} & 0\\ \hline
	$d-1$ & 0 & 0 & 0 & 0 & $\cdots$ & 0 & 0 & $c_{d-1}^2$\\
	\hline
\end{tabular}
\end{center}
\end{table}


\begin{table}[H]
\caption{$2\times 2$ block correlation table $C_{x=0,y=0,m}$ and $C_{x=0,y=1,m}$}
\label{tab:corr1}
\begin{center}
\begin{tabular}{| c || c | c |}
	\hline
	$a \backslash b$ & 2m & 2m+1\\ \hline \hline
	2m & $c_{2m}^2\cos^2{(\frac{\mu_m}{2})}$ & $c_{2m}^2\sin^2{(\frac{\mu_m}{2})}$ \\ \hline
	2m+1 & $c_{2m+1}^2\sin^2{(\frac{\mu_m}{2})}$ & $c_{2m+1}^2\cos^2{(\frac{\mu_m}{2})}$  \\
	\hline
\end{tabular}
\end{center}
\end{table}

\begin{table}[H]
\caption{$2 \times 2$ block correlation table $C_{x=1,y=0,m}$}
\label{tab:corr3}
\begin{center}
\begin{tabular}{| c || c | c |}
	\hline
	$a \backslash b$ & 2m & 2m+1\\ \hline \hline
	2m & $\frac{1}{2}(c_{2m}\cos{(\frac{\mu_m}{2})}+c_{2m+1}\sin{(\frac{\mu_m}{2})})^2$ & $\frac{1}{2}(c_{2m+1}\cos{(\frac{\mu_m}{2})}-c_{2m}\sin{(\frac{\mu_m}{2})})^2$ \\ \hline
	2m+1 & $\frac{1}{2}(c_{2m}\cos{(\frac{\mu_m}{2})}-c_{2m+1}\sin{(\frac{\mu_m}{2})})^2$ & $\frac{1}{2}(c_{2m+1}\cos{(\frac{\mu_m}{2})}+c_{2m}\sin{(\frac{\mu_m}{2})})^2$  \\
	\hline
\end{tabular}
\end{center}
\end{table}

\begin{table}[H]
\caption{$2 \times 2$ block correlation table $C_{x=1,y=1,m}$}
\label{tab:corr4}
\begin{center}
\begin{tabular}{| c || c | c |}
	\hline
	$a \backslash b$ & 2m & 2m+1\\ \hline \hline
	2m & $\frac{1}{2}(c_{2m}\cos{(\frac{\mu_m}{2})}-c_{2m+1}\sin{(\frac{\mu_m}{2})})^2$ & $\frac{1}{2}(c_{2m+1}\cos{(\frac{\mu_m}{2})}+c_{2m}\sin{(\frac{\mu_m}{2})})^2$ \\ \hline
	2m+1 & $\frac{1}{2}(c_{2m}\cos{(\frac{\mu_m}{2})}+c_{2m+1}\sin{(\frac{\mu_m}{2})})^2$ & $\frac{1}{2}(c_{2m+1}\cos{(\frac{\mu_m}{2})}-c_{2m}\sin{(\frac{\mu_m}{2})})^2$  \\
	\hline
\end{tabular}
\end{center}
\end{table}

\item[(ii)] Similarly, for measurement settings $x \in \{0,2\}$ and $y \in \{2,3\}$ the correlation tables $T_{x,y}$ are also block-diagonal, but ``shifted down'' appropriately by one measurement outcome. The $2\times 2$ blocks are $D_{x,y,m}$ (corresponding to outcomes $2m+1$ and $2m+2$) for $x \in \{0,2\}$ and $y \in \{2,3\}$, defined as $D_{x,y,m}:= (c_{2m+1}^2+c_{2m+2}^2) \cdot C^{ideal}_{f(x),g(y); \theta_m'}$, where $\theta_m' :=\arctan\big(\frac{c_{2m+2}}{c_{2m+1}}\big) \in (0,\frac{\pi}{2})$, and $f(0) = 0, f(2) = 1, g(2) = 0, g(3) = 1$. The correlations, $T_{x,y}$, for $x \in \{0,2\}$ and $y \in \{2,3\}$ are given precisely in Tables~\ref{tab:txy3} to \ref{tab:corr8} where $\mu'_{m} := \arctan(\sin(2\theta'_m))$.

\begin{table}[H]
\caption{$T_{x,y}$ for $x\in \{0,2\}$ and $y\in \{2,3\}$, for even values of $d \geq 2$}
\label{tab:txy3}
\begin{center}
\begin{tabular}{| c || c | c | c | c | c | c | c |}
	\hline
	$a \backslash b$ & 1 & 2 & 3 & 4 & $\cdots$ & $d-1$ & 0 \\ \hline \hline
	1 & \multicolumn{2}{| c |}{\multirow{2}{*}{$D_{x,y,m=0}$}} & 0 & 0 & $\cdots$ & 0 & 0 \\ 
	\hhline{*{1}{|-}*{2}{|~}*{5}{|-}}
	2 & \multicolumn{2}{| c |}{} & 0 & 0 & $\cdots$ & 0 & 0 \\ \hline
	3 & 0 & 0 & \multicolumn{2}{| c |}{\multirow{2}{*}{$D_{x,y,m=1}$}} & $\cdots$ & 0 & 0 \\
	\hhline{*{3}{|-}*{2}{|~}*{3}{|-}}
	4 & 0 & 0 & \multicolumn{2}{| c |}{} & $\cdots$ & 0 & 0 \\ \hline
	$\vdots$ & $\vdots$ & $\vdots$ & $\vdots$ & $\vdots$ & $\ddots$ & $\vdots$ & $\vdots$  \\ \hline
	$d-1$ & 0 & 0 & 0 & 0 & $\cdots$ &  \multicolumn{2}{| c |}{\multirow{2}{*}{$D_{x,y,m=\frac{d}{2}-1}$}} \\
	\hhline{*{6}{|-}*{2}{|~}}
	0 & 0 & 0 & 0 & 0 & $\cdots$ & \multicolumn{2}{| c |}{} \\ 
	\hline
\end{tabular}
\end{center}
\end{table}
\begin{table}[H]
\caption{$T_{x,y}$ for $x\in \{0,2\}$ and $y\in \{2,3\}$, for odd values of $d \geq 3$}
\label{tab:txy4}
\begin{center}
\begin{tabular}{| c || c | c | c | c | c | c | c | c |}
	\hline
	$a \backslash b$ & 1 & 2 & 3 & 4 & $\cdots$ & $d-2$ & $d-1$ & 0 \\ \hline \hline
	1 & \multicolumn{2}{| c |}{\multirow{2}{*}{$D_{x,y,m=0}$}} & 0 & 0 & $\cdots$ & 0 & 0 & 0\\ 
	\hhline{*{1}{|-}*{2}{|~}*{6}{|-}}
	2 & \multicolumn{2}{| c |}{} & 0 & 0 & $\cdots$ & 0 & 0 & 0\\ \hline
	3 & 0 & 0 & \multicolumn{2}{| c |}{\multirow{2}{*}{$D_{x,y,m=1}$}} & $\cdots$ & 0 & 0 & 0\\
	\hhline{*{3}{|-}*{2}{|~}*{4}{|-}}
	4 & 0 & 0 & \multicolumn{2}{| c |}{} & $\cdots$ & 0 & 0 & 0\\ \hline
	$\vdots$ & $\vdots$ & $\vdots$ & $\vdots$ & $\vdots$ & $\ddots$ & $\vdots$ & $\vdots$  & 0\\ \hline
	$d-2$ & 0 & 0 & 0 & 0 & $\cdots$ &  \multicolumn{2}{| c |}{\multirow{2}{*}{$D_{x,y,m=\frac{d-3}{2}}$}} & 0\\
	\hhline{*{6}{|-}*{2}{|~}*{1}{|-}}
	$d-1$ & 0 & 0 & 0 & 0 & $\cdots$ & \multicolumn{2}{| c |}{} & 0\\ \hline
	0 & 0 & 0 & 0 & 0 & $\cdots$ & 0 & 0 & $c_0^2$\\
	\hline
\end{tabular}
\end{center}
\end{table}

\begin{table}[H]
\caption{$2 \times 2$ block correlation table $D_{x=0,y=2,m}$ and $D_{x=0,y=3,m}$}
\label{tab:corr5}
\begin{center}
\begin{tabular}{| c || c | c |}
	\hline
	$a \backslash b$ & 2m+1 & 2m+2\\ \hline \hline
	2m+1 & $c_{2m+1}^2\cos^2{(\frac{\mu'_{m}}{2})}$ & $c_{2m+1}^2\sin^2{(\frac{\mu'_{m}}{2})}$ \\ \hline
	2m+2 & $c_{2m+2}^2\sin^2{(\frac{\mu'_{m}}{2})}$ & $c_{2m+2}^2\cos^2{(\frac{\mu'_{m}}{2})}$  \\
	\hline
\end{tabular}
\end{center}
\end{table}

\begin{table}[H]
\caption{$2 \times 2$ block correlation table $D_{x=2,y=2,m}$}
\label{tab:corr7}
\begin{center}
\begin{tabular}{| c || c | c |}
	\hline
	$a \backslash b$ & 2m+1 & 2m+2\\ \hline \hline
	2m+1 & $\frac{1}{2}(c_{2m+1}\cos{(\frac{\mu'_{m}}{2})}+c_{2m+2}\sin{(\frac{\mu'_{m}}{2})})^2$ & $\frac{1}{2}(c_{2m+2}\cos{(\frac{\mu'_{m}}{2})}-c_{2m+1}\sin{(\frac{\mu'_{m}}{2})})^2$ \\ \hline
	2m+2 & $\frac{1}{2}(c_{2m+1}\cos{(\frac{\mu'_{m}}{2})}-c_{2m+2}\sin{(\frac{\mu'_{m}}{2})})^2$ & $\frac{1}{2}(c_{2m+2}\cos{(\frac{\mu'_{m}}{2})}+c_{2m+1}\sin{(\frac{\mu'_{m}}{2})})^2$  \\
	\hline
\end{tabular}
\end{center}
\end{table}

\begin{table}[H]
\caption{$2 \times 2$ block correlation table $D_{x=2,y=3,m}$}
\label{tab:corr8}
\begin{center}
\begin{tabular}{| c || c | c |}
	\hline
	$a \backslash b$ & 2m+1 & 2m+2\\ \hline \hline
	2m+1 & $\frac{1}{2}(c_{2m+1}\cos{(\frac{\mu'_{m}}{2})}-c_{2m+2}\sin{(\frac{\mu'_{m}}{2})})^2$ & $\frac{1}{2}(c_{2m+2}\cos{(\frac{\mu'_{m}}{2})}+c_{2m+1}\sin{(\frac{\mu'_{m}}{2})})^2$ \\ \hline
	2m+2 & $\frac{1}{2}(c_{2m+1}\cos{(\frac{\mu'_{m}}{2})}+c_{2m+2}\sin{(\frac{\mu'_{m}}{2})})^2$ & $\frac{1}{2}(c_{2m+2}\cos{(\frac{\mu'_{m}}{2})}-c_{2m+1}\sin{(\frac{\mu'_{m}}{2})})^2$  \\
	\hline
\end{tabular}
\end{center}
\end{table}

\end{enumerate}

\subsection{The ideal measurements}
\label{ideal measurements}
We now explicitly provide the ideal measurements on $\ket{\psi_{\text{target}}}=\sum_{i=0}^{d-1}c_i\ket{ii}$ that satisfy the constraints described above, and we refer to the correlation produced by the ideal measurements as the \textit{ideal correlation}.

Let $\sigma_Z$ and $\sigma_X$ be the usual Pauli matrices. For a single-qubit observable $A$, we denote by $[A]_m$ the observable defined with respect to the basis $\{\ket{2m\mod d},\ket{(2m+1) \mod d}\}$. For example, $[\sigma_Z]_m = \ket{2m}\bra{2m} - \ket{2m+1}\bra{2m+1}$. Similarly, we denote by $[A]'_m$ the observable defined with respect to the basis $\{\ket{(2m+1)\mod d}, \ket{(2m+2)\mod d}\}$. We use the notation $\bigoplus A_i$ to denote the direct sum of observables $A_i$.

For $x=0$: Alice measures in the computational basis (i.e. in the basis $\{\ket{0},\ket{1},\cdots,\ket{d-1}\}$). For $x=1$ and $x=2$: for $d$ even, she measures in the eigenbases of observables $\bigoplus_{m=0}^{\frac{d}{2}-1} [\sigma_X]_m$ and $\bigoplus_{m=0}^{\frac{d}{2}-1} [\sigma_X]'_m$ respectively, with the natural assignments of $d$ measurement outcomes; for $d$ odd, she measures in the eigenbases of observables $\bigoplus_{m=0}^{\frac{d-1}{2}-1} [\sigma_X]_m\oplus \ket{d-1}\bra{d-1}$ and $\ket{0}\bra{0}\oplus\bigoplus_{m=0}^{\frac{d-1}{2}-1} [\sigma_X]'_m$ respectively.

In a similar way, for $y=0$ and $y=1$: for $d$ even, Bob measures in the eigenbases of $\bigoplus_{m=0}^{\frac{d}{2}-1} [\cos{(\mu_m)}\sigma_Z+\sin{(\mu_m)}\sigma_X]_m$ and $\bigoplus_{m=0}^{\frac{d}{2}-1} [\cos{(\mu_m)}\sigma_Z-\sin{(\mu_m)}\sigma_X]_m$ respectively, with the natural assignments of $d$ measurement outcomes, where here $\mu_m = \arctan(\sin(2\theta_m))$ and $\theta_m  = \arctan(\frac{c_{2m+1}}{c_{2m}})$; for $d$ odd, he measures in the eigenbases of $\bigoplus_{m=0}^{\frac{d-1}{2}-1} [\cos{(\mu_m)}\sigma_Z+\sin{(\mu_m)}\sigma_X]_m\oplus\ket{d-1}\bra{d-1}$ and $\bigoplus_{m=0}^{\frac{d-1}{2}-1} [\cos{(\mu_m)}\sigma_Z-\sin{(\mu_m)}\sigma_X]_m\oplus\ket{d-1}\bra{d-1}$ respectively.

For $y=2$ and $y=3$: for $d$ even, Bob measures in the eigenbases of $\bigoplus_{m=0}^{\frac{d}{2}-1} [\cos{(\mu'_m)}\sigma_Z+\sin{(\mu'_m)}\sigma_X]'_m$ and $\bigoplus_{m=0}^{\frac{d}{2}-1} [\cos{(\mu'_m)}\sigma_Z-\sin{(\mu'_m)}\sigma_X]'_m$ respectively, where $\mu'_m = \arctan(\sin(2\theta'_m))$ and $\theta'_m = \arctan(\frac{c_{2m+2}}{c_{2m+1}})$; for $d$ odd, he measures in the eigenbases of $\ket{0}\bra{0}\oplus\bigoplus_{m=0}^{\frac{d-1}{2}-1} [\cos{(\mu'_m)}\sigma_Z+\sin{(\mu'_m)}\sigma_X]'_m$ and $\ket{0}\bra{0}\oplus\bigoplus_{m=0}^{\frac{d-1}{2}-1} [\cos{(\mu'_m)}\sigma_Z-\sin{(\mu'_m)}\sigma_X]'_m$ respectively.



\section{Proof of self-testing}
\label{proof}
This section is dedicated entirely to proving Theorem \ref{main theorem}. Most of the work in the proof is aimed at constructing operators satisfying the sufficient conditions from Lemma \ref{YNcriterion}. This, explicitly, means constructing appropriate projections $P_A^{(k)}, P_B^{(k)}$ and unitaries $X_A^{(k)}, X_B^{(k)}$. In subsection \ref{consequences}, we construct the projections, and, moreover, certain unitary ``flip'' operators $X^{\mathscr{u}}_{A,m}, X^{'\mathscr{u}}_{A,m}$. In subsection \ref{connecting}, we show how to obtain unitaries $X_A^{(k)}, X_B^{(k)}$ as appropriate alternating products of the flip operators.  Finally, we argue that the same local isometry given by Lemma \ref{YNcriterion} works also to self-test the ideal measurements from subsection \ref{ideal measurements}.

\subsection{Constructing the projections and the ``flip'' operators}
\label{consequences}

Recall that we denote by $\Pi^{A_x}_{i}$ the projection corresponding to Alice obtaining outcome $i$ on measurement setting $x$, and similarly for the $\Pi^{B_y}_{i}$ on Bob's side. We will first derive consequences that follow from the constraints in item (i) of subsection \ref{correlations}, that we imposed on our correlations. The constraints in item (ii) of subsection \ref{correlations} have similar implications.

We define the operators $\hat{A}_{x,m}=\Pi^{A_x}_{2m}-\Pi^{A_x}_{2m+1}$ and $\hat{B}_{y,m}=\Pi^{B_y}_{2m}-\Pi^{B_y}_{2m+1}$ for $x,y \in \{0,1\}$. Clearly, $(\Ah_{x,m})^2 = \Pi^{A_x}_{2m} + \Pi^{A_x}_{2m+1} := \mathds{1}_m^{A_x}$ and $(\Bh_{y,m})^2 = \Pi^{B_y}_{2m} + \Pi^{B_y}_{2m+1} := \mathds{1}_m^{B_y}$.\\
Now, $\|\Pi^{A_0}_{2m} \ps\| = \sqrt{\average{\psi| \Pi^{A_0}_{2m}| \psi}} = \sqrt{\average{\psi| \Pi^{A_0}_{2m} \cdot \sum_{i=0}^{d-1}\Pi^{B_0}_{i} |\psi}} = \sqrt{c_{2m}^2\cos^2{(\frac{\mu_m}{2})} + c_{2m}^2\sin^2{(\frac{\mu_m}{2})}} = c_{2m}$, and $\|\Pi^{A_0}_{2m+1} \ps\| =  c_{2m+1}$. With similar other calculations we deduce that
\begin{equation}
\label{id_norms}
\|\mathds{1}_m^{A_i} \ps \| = \|\mathds{1}_m^{B_j} \ps\| = \sqrt{c_{2m}^2 + c_{2m+1}^2}\,\,\,\, \forall i,j \in \{0,1\}\,.
\end{equation}
Moreover, notice that $\average{\psi| \mathds{1}_m^{A_i}\mathds{1}_m^{B_j}| \psi} = c_{2m}^2 + c_{2m+1}^2 =  \|\mathds{1}_m^{A_i} \ps \| \cdot \|\mathds{1}_m^{B_j} \ps \|$. Hence, by Cauchy-Schwarz, it must be the case that \begin{equation}
\label{id_equalities}
\mathds{1}_m^{A_i} \ps = \mathds{1}_m^{B_j} \ps \,\,\,\, \forall i,j \in \{0,1\}\,.
\end{equation}

\noindent By design, the correlations are such that 
\begin{equation}
\bra{\psi}\alpha_m\hat{A}_{0,m}+\hat{A}_{0,m}\hat{B}_{0,m}+\hat{A}_{0,m}\hat{B}_{1,m}+\hat{A}_{1,m}\hat{B}_{0,m}-\hat{A}_{1,m}\hat{B}_{1,m}\ket{\psi}=\sqrt{8+2\alpha_m^2} \cdot (c_{2m}^2 + c_{2m+1}^2) \label{almost_tilted_chsh}
\end{equation}
where $\alpha_m=\frac{2}{\sqrt{1+2\tan^2{(2\theta_m)}}}$. As such, this is not a maximal violation of the tilted CHSH inequality (since $\ps$ has unit norm). However, we can get around this by defining the normalised state $\ket{\psi_m}=\frac{\Id^{A_0}_m\ket{\psi}}{\sqrt{c_{2m}^2+c_{2m+1}^2}}$. Since $\Ah_{i,m} \ps = \Ah_{i,m} \mathds{1}_m^{A_i} \ps = \Ah_{i,m} \mathds{1}_m^{A_0} \ps$, and $\Bh_{i,m} \ps = \Bh_{i,m} \mathds{1}_m^{B_i} \ps = \Bh_{i,m} \mathds{1}_m^{A_0} \ps$, by (\ref{id_equalities}), then (\ref{almost_tilted_chsh}) implies
=\begin{equation}
\label{eq12}
\bra{\psi_m}\alpha_m\hat{A}_{0,m}+\hat{A}_{0,m}\hat{B}_{0,m}+\hat{A}_{0,m}\hat{B}_{1,m}+\hat{A}_{1,m}\hat{B}_{0,m}-\hat{A}_{1,m}\hat{B}_{1,m}\ket{\psi_m}=\sqrt{8+2\alpha_m^2}
\end{equation}
Now, define the ``unitarized'' versions of the operators in \eqref{eq12}: $\hat{A}^{\mathscr{u}}_{i,m} := \mathds{1}-\mathds{1}_m^{A_i} + \hat{A}_{i,m}$ and $\hat{B}^{\mathscr{u}}_{i,m} := \mathds{1}-\mathds{1}_m^{B_i} + \hat{B}_{i,m}$. Then clearly equation \eqref{eq12} holds also with the unitarized operators, by definition of $\ket{\psi_m}$. Now, let $Z^{\mathscr{u}}_{A,m} := \hat{A}^{\mathscr{u}}_{0,m}$, $X^{\mathscr{u}}_{A,m} := \Ah^{\mathscr{u}}_{1,m}$. Then, let $Z^*_{B,m}$ and $X^*_{B,m}$ be respectively $\frac{\Bh^{\mathscr{u}}_{0,m}+ \Bh^{\mathscr{u}}_{1,m}}{2\cos(\mu_m)}$ and $\frac{\Bh^{\mathscr{u}}_{0,m}- \Bh^{\mathscr{u}}_{1,m}}{2\sin(\mu_m)}$, but with all zero eigenvalues replaced by one, and define $Z^{\mathscr{u}}_{B,m} = Z^*_{B,m}|Z^*_{B,m}|^{-1}$ and $X^{\mathscr{u}}_{B,m} = X^*_{B,m}|X^*_{B,m}|^{-1}$ (this is again a required unitarization step). Then, by Lemma \ref{Bamps lemma}, the above maximal violation of the tilted CHSH inequality implies that
\begin{align}
Z^{\mathscr{u}}_{A,m} \ket{\psi_m} &= Z^{\mathscr{u}}_{B,m} \ket{\psi_m} \label{44}\\
X^{\mathscr{u}}_{A,m} (\mathds{1} - Z^{\mathscr{u}}_{A,m}) \ket{\psi_m} &= \tan(\theta_m) X^{\mathscr{u}}_{B,m} (\mathds{1} + Z^{\mathscr{u}}_{A,m}) \ket{\psi_m} \label{52}
\end{align}

Define the subspace $\mathcal{B}_m=\mbox{range}(\mathds{1}_m^{B_0}) + \mbox{range}(\mathds{1}_m^{B_1})$, and the projection $\mathds{1}_{\mathcal{B}_m}$ onto subspace $\mathcal{B}_m$. Then, notice from the way $Z^{\mathscr{u}}_{B,m}$ is defined, that it can be written as $Z^{\mathscr{u}}_{B,m} = \mathds{1}-\mathds{1}_{\mathcal{B}_m} + \tilde{Z}_{B,m}$, where $\tilde{Z}_{B,m}$ is some operator living entirely on subspace $\mathcal{B}_m$. This implies that $Z^{\mathscr{u}}_{B,m} \ket{\psi_m} =  \tilde{Z}_{B,m} \ket{\psi_m} = \tilde{Z}_{B,m} \ps$, where we have used \eqref{id_equalities} and the fact that 
\begin{eqnarray}\label{onesb}
\mathds{1}_m^{B_0} \ket{\psi} = \mathds{1}_m^{B_1} \ket{\psi}&\Longrightarrow& \mathds{1}_{\mathcal{B}_m} \ket{\psi} = \mathds{1}_m^{B_i}\ket{\psi}\,.
\end{eqnarray}

Hence, from \eqref{44} we deduce that $\hat{A}_{0,m} \ps= \tilde{Z}_{B,m} \ps$. Define projections $P_A^{(2m)} := (\mathds{1}_m^{A_0} + \hat{A}_{0,m})/2 = \Pi_{2m}^{A_0}$, $P_A^{(2m+1)} := (\mathds{1}_m^{A_0} - \hat{A}_{0,m})/2 = \Pi_{2m+1}^{A_0}$, $P_B^{(2m)} := (\mathds{1}_{\mathcal{B}_m} + \tilde{Z}_{B,m})/2$ and $P_B^{(2m+1)} := (\mathds{1}_{\mathcal{B}_m} - \tilde{Z}_{B,m})/2$. 

Note that $P_B^{(2m)},P_B^{(2m+1)}$ are indeed projections, since $\tilde{Z}_{B,m}$ has all $\pm 1$ eigenvalues corresponding to subspace $\mathcal{B}_m$, and is zero outside. We also have, for all $m$ and $k=2m,2m+1$,
\begin{align}
\label{eq16}
P_A^{(k)} \ps = (\mathds{1}_m^{A_0} +(-1)^k \hat{A}_{0,m})/2 \ps &=  (\mathds{1}_m^{B_0} +(-1)^k \hat{A}_{0,m})/2 \ps \nonumber\\
&= (\mathds{1}_{\mathcal{B}_m} +(-1)^k \tilde{Z}_{B,m})/2 \ps = P_B^{(k)} \ps
\end{align}
Further, notice that $(\mathds{1} +(-1)^k Z^{\mathscr{u}}_{A,m}) \ket{\psi_m} =  (\mathds{1}_m^{A_0} +(-1)^k\hat{A}_{0,m}) \ket{\psi_m} = (\mathds{1}_m^{A_0} +(-1)^k \hat{A}_{0,m}) \ps = P_A^{(k)}\ps$. Plugging this into \eqref{52}, gives
\begin{equation}
\label{eqimp}
X^{\mathscr{u}}_{A,m} P_A^{(2m+1)} \ps = \tan(\theta_m) X^{\mathscr{u}}_{B,m} P_A^{(2m)} \ps  = \frac{c_{2m+1}}{c_{2m}} X^{\mathscr{u}}_{B,m} P_A^{(2m)} \ps\\
\end{equation}

\vspace{5mm}

Now, we turn to the constraints on our correlations that we imposed in item (ii) of subsection \ref{correlations}. These have similar implications to the ones we just derived. 

We can similarly define the operators $\hat{A}'_{0,m}=\Pi^{A_0}_{2m+1}-\Pi^{A_0}_{2m+2}$, $\hat{A}'_{1,m}=\Pi^{A_2}_{2m+1}-\Pi^{A_2}_{2m+2}$ $\hat{B}'_{0,m}=\Pi^{B_2}_{2m+1}-\Pi^{B_2}_{2m+2}$,
$\hat{B}'_{1,m}=\Pi^{B_3}_{2m+1}-\Pi^{B_3}_{2m+2}$, and
$\Id_m^{A_x'}=\left(\hat{A}'_{x,m}\right)^2$ and $\Id_m^{B_y'}=\left(\hat{B}'_{y,m}\right)^2$. Using the argument employed earlier and following the same procedure, we can analogously construct unitary operators $Z^{'\mathscr{u}}_{A,m}$, $X^{'\mathscr{u}}_{A,m}$, $Z^{'\mathscr{u}}_{B,m}$ and $X^{'\mathscr{u}}_{B,m}$ from operators $\hat{A}'_{x,m}$ and $\hat{B}'_{y,m}$.

\begin{align}
Z'_{A,m} \ket{\psi'_m} &= Z'_{B,m} \ket{\psi'_m} \\
X'_{A,m} (\mathds{1}_m^{A'_0} - Z'_{A,m}) \ket{\psi'_m} &= \tan(\theta'_m) X'_{B,m} (\mathds{1}_m^{A'_0} + Z'_{A,m})\ket{\psi'_m} 
\end{align}

where $\ket{\psi'_m} = \frac{\Id_m^{A_0'}\ps}{\sqrt{c_{2m+1}^2+c_{2m+2}^2}}$. And from here, with the same steps as above, we deduce that 
\begin{equation}
\label{eq20}
X^{'\mathscr{u}}_{A,m} P_A^{(2m+2)} \ps = \tan(\theta'_m) X^{'\mathscr{u}}_{B,m} P_A^{(2m+1)} \ps = \frac{c_{2m+2}}{c_{2m+1}} X^{'\mathscr{u}}_{B,m} P_A^{(2m+1)} \ps \\
\end{equation}

\subsection{Constructing the unitaries}
\label{connecting}
For notational convenience, we drop the superscript $\mathscr{u}$ from the unitary operators $X^{\mathscr{u}}_{A/B,m}$,$X^{'\mathscr{u}}_{A/B,m}$ in equations \eqref{52} and \eqref{eq20} of the previous subsection. We also rename $X^{'\mathscr{u}}_{A/B,m}$ as $Y_{A/B,m}$. Then, we recall equations \eqref{52} and \eqref{eq20}:

\begin{align}
X_{A,m} P_A^{2m+1} \ps &= \frac{c_{2m+1}}{c_{2m}} X_{B,m} P_A^{2m} \ps \label{eq21}\\
Y_{A,m} P_A^{2m+2} \ps &= \frac{c_{2m+2}}{c_{2m+1}} Y_{B,m} P_A^{2m+1} \label{eq22}\ps
\end{align}

Recall that we ultimately wish to produce unitary operators satisfying condition \eqref{c3} from Lemma \ref{YNcriterion}. The operators $X_{A/B,m}$ and $Y_{A/B,m}$ can be can be intuitively thought of as ``flip operators'', in the sense that $X_{A,m}$ acts on $P^{(2m+1)}_{A}\ket{\psi}$ (which is equal to $P^{(2m+1)}_{B}\ket{\psi}$ when condition \eqref{c2} is satisfied) and turns it into $X_{B,m}P^{(2m)}_{A}\ket{\psi}$, up to an appropriate factor. On the other hand, the flip operator $Y_{A,m}$ will turn $P^{(2m)}_{A}\ket{\psi}$ into $Y_{B,m}P^{(2m-1)}_{A}\ket{\psi}$, up to a factor. The idea is, then, that the appropriate alternating product of these unitary flip operators will turn $P^{(i)}_{A}\ket{\psi}$ into precisely $\frac{c_i}{c_0}(X^{(i)}_{B})^\dagger P^{(0)}_{A}\ket{\psi}$, which is the behaviour required from condition \eqref{c3} of Lemma \ref{YNcriterion}, when we let these alternating products be the $X^{(i)}_{A}$ and $X^{(i)}_{B}$ from \eqref{c3}.

We have already shown, in \eqref{eq16}, that the $P_{A/B}^{(k)}$, as defined in the previous subsection, satisfy $P_A^{(k)} \ps = P_B^{(k)} \ps$ for $k=0,..,d-1$, i.e. condition \eqref{c2} from Lemma \ref{YNcriterion}, with the $P^{(k)}_A$ forming, by definition, a complete set of orthogonal projections. 

We are ready to define $X_{A/B}^{(k)}$ as follows:
\begin{equation}
X_{A}^{(k)} =  
\begin{cases}
    \mathds{1},    & \text{if } k=0\\
    X_{A,0}Y_{A,0}X_{A,1}Y_{A,1} \ldots X_{A,m-1}Y_{A,m-1}X_{A,m}   & \text{if } k=2m+1\\
    X_{A,0}Y_{A,0}X_{A,1}Y_{A,1} \ldots X_{A,m-1}Y_{A,m-1},              & \text{if } k=2m
\end{cases}
\end{equation}
and 
\begin{equation}
X_{B}^{(k)} =  
\begin{cases}
    \mathds{1},    & \text{if } k=0\\
    X_{B,0}Y_{B,0}X_{B,1}Y_{B,1} \ldots X_{B,m-1}Y_{B,m-1}X_{B,m}   & \text{if } k=2m+1\\
    X_{B,0}Y_{B,0}X_{B,1}Y_{B,1} \ldots X_{B,m-1}Y_{B,m-1},              & \text{if } k=2m
\end{cases}
\end{equation}
Note that $X_{A}^{(k)}$ and $X_{B}^{(k)}$ are unitary since they are product of unitaries.
Finally we check that condition \eqref{c3} holds, namely 
\begin{equation}\label{eqfinal}
X_A^{(k)} P_A^{(k)} \ps = \frac{c_k}{c_0}(X_B^{(k)})^{\dagger} P_A^{(0)} \ps
\end{equation}
The case $k=0$,
\begin{align}
    X^{(0)}_A P^{(0)}_A \ket{\psi} &= \Id P^{(0)}_A \ket{\psi}\nonumber\\
    &= \frac{c_0}{c_0} X^{(0)}_B P^{(0)}_A \ket{\psi}.
\end{align}
For $k=2m+1$, 
\begingroup
\allowdisplaybreaks
\begin{align}
X_A^{(k)} P_A^{(k)}\ps &= X_{A,0}Y_{A,0}X_{A,1}Y_{A,1} \ldots X_{A,m-1}Y_{A,m-1}X_{A,m} P_A^{(2m+1)} \ps \nonumber\\
&\stackrel{\eqref{eq21}}{=}\frac{c_{2m+1}}{c_{2m}}X_{A,0}Y_{A,0}X_{A,1}Y_{A,1} \ldots X_{A,m-1}Y_{A,m-1} X_{B,m} P_A^{(2m)} \ps
\nonumber\\
&= \frac{c_{2m+1}}{c_{2m}}X_{B,m} X_{A,0}Y_{A,0}X_{A,1}Y_{A,1} \ldots X_{A,m-1}Y_{A,m-1}P_A^{(2m)} \ps \nonumber\\
&\stackrel{\eqref{eq22}}{=} \frac{c_{2m+1}}{c_{2m}}\cdot \frac{c_{2m}}{c_{2m-1}}X_{B,m} X_{A,0}Y_{A,0}X_{A,1}Y_{A,1} \ldots X_{A,m-1} Y_{B,m-1} P_A^{(2m-1)}\ps
\nonumber\\
&= \frac{c_{2m+1}}{c_{2m}}\cdot \frac{c_{2m}}{c_{2m-1}}X_{B,m}Y_{B,m-1}X_{A,0}Y_{A,0}X_{A,1}Y_{A,1}\ldots X_{A,m-1}P_A^{(2m-1)} \ps  \nonumber\\
&= \ldots  \nonumber\\
&= \frac{c_{2m+1}}{\cancel{c_{2m}}} \cdot \frac{\cancel{c_{2m}}}{\cancel{c_{2m-1}}} \ldots \frac{\cancel{c_2}}{\cancel{c_1}} \cdot \frac{\cancel{c_1}}{c_0} X_{B,m}Y_{B,m-1}X_{B,m-1} \dots Y_{B,1}X_{B,1}Y_{B,0}X_{B,0} P_A^{(0)} \ps \nonumber\\
&= \frac{c_{2m+1}}{c_0}(X_B^{(k)})^{\dagger} P_A^{(0)}\ps \label{eq31} 
\end{align}
\endgroup
which is indeed \eqref{eqfinal}, as $2m+1=k$. The case $k=2m$ is treated similarly. This completes the construction of the local isometry $\Phi$, by Lemma \ref{YNcriterion}. To conclude the proof of Theorem \ref{main theorem}, we just need to show that this isometry also self-tests the ideal measurements given precisely below. The rest of the proof is included in Appendix \ref{self-testing measurements}.

We emphasize that the whole proof goes through in the same way if we replace $\ps$ with a general mixed state. In particular, one simply replaces all equalities between vectors with equalities between density matrices. Moreover, the Euclidean inner product is replaced by $\langle\cdot,\cdot\rangle: \mathcal{L}(\text{supp}\rho, \mathcal{H}_A \otimes \mathcal{H}_B) \times \mathcal{L}(\text{supp}\rho, \mathcal{H}_A \otimes \mathcal{H}_B) \rightarrow \mathbb{C}$
such that 
\begin{equation}
\langle A,B\rangle := Tr (AB^{\dagger}\rho),
\end{equation}
where $\text{supp}\rho = \{\ket{\phi} \in \mathcal{H} : \rho \ket{\phi} \neq 0\}$, and $\mathcal{L}(\text{supp}\rho, \mathcal{H}_A \otimes \mathcal{H}_B)$ is the space of linear maps from $\text{supp}\rho$ to $\mathcal{H}_A \otimes \mathcal{H}_B$. Notice that the product defined above doesn't in general satisfy the symmetric property of inner products. Nonetheless, Cauchy-Schwarz still holds on instances that satisfy the symmetry property (in particular when $A$ and $B$ commute). So, as an example, we would replace the expression $\average{\psi| \mathds{1}_m^{A_i}\mathds{1}_m^{B_j}| \psi}$, after equation \eqref{id_norms}, with $\langle \mathds{1}_m^{A_i}|_{\text{supp}\rho} ,\mathds{1}_m^{B_j}|_{\text{supp}\rho}\rangle = Tr (\mathds{1}_m^{A_i}|_{\text{supp}\rho} \mathds{1}_m^{B_j}|_{\text{supp}\rho}\,\rho)$, and deduce, through Cauchy-Schwarz, that $\mathds{1}_m^{A_i}|_{\text{supp}\rho} = \mathds{1}_m^{B_j}|_{\text{supp}\rho}$.

Finally, Lemma \ref{Bamps lemma}, from Bamps and Pironio \cite{Bamps2015}, as well as Lemma \ref{YNcriterion}, hold analogously in corresponding mixed state form.


\section{Discussion}
\label{discussion}

\begin{figure}[H]
  \centering  \includegraphics[width=0.55\textwidth]{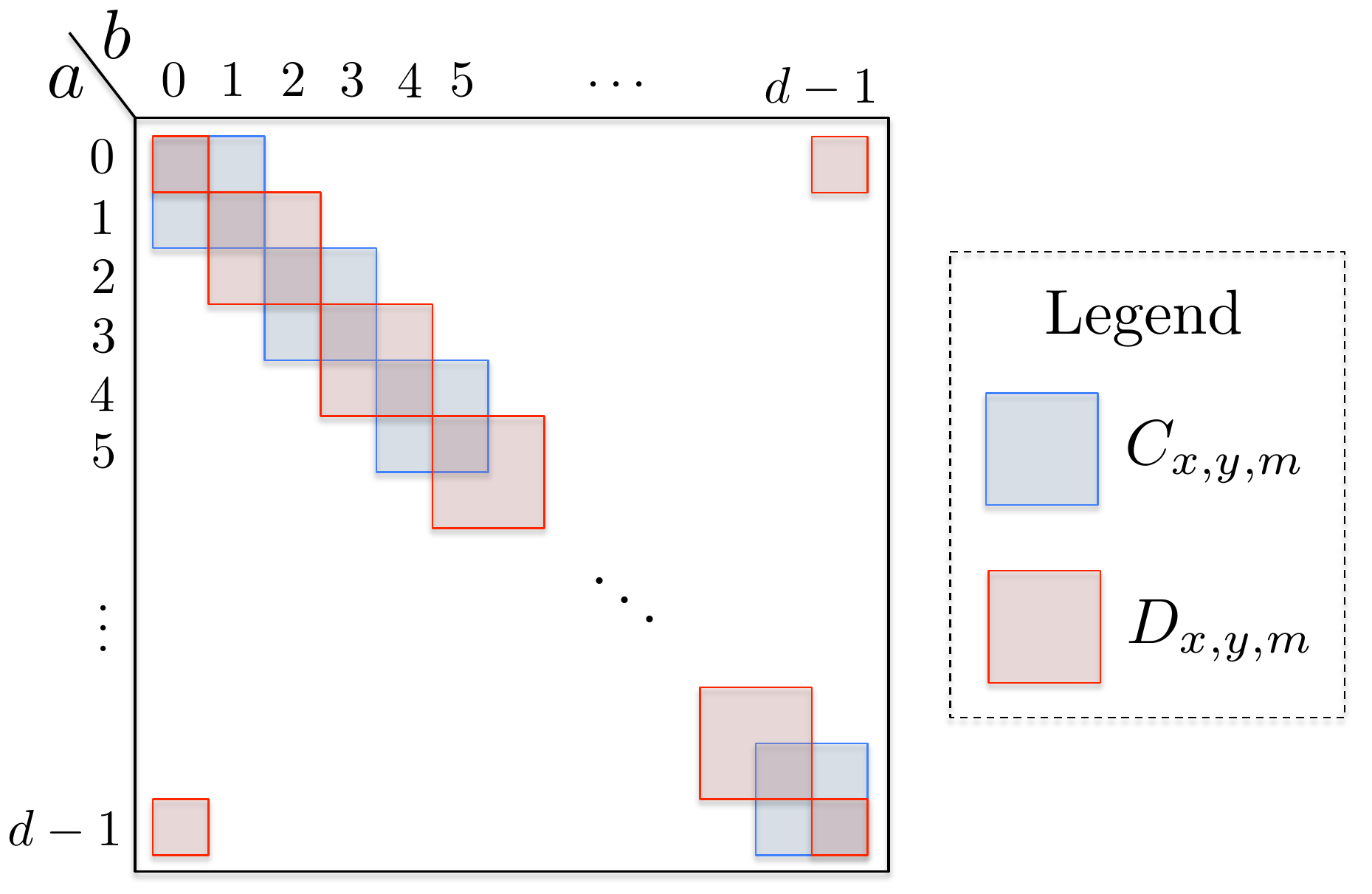}
      \caption{Block-diagonal structure of the correlation tables}
      \label{fig:framework}
\end{figure}

In our proof, we described explicit self-testing correlations for the $2\times 2$ blocks, in Tables \ref{tab:corr1}-\ref{tab:corr4} and \ref{tab:corr5}-\ref{tab:corr8}. However, we remark that this is not the only choice of correlations that can be made to self-test all bipartite entangled states. In fact, as a natural consequence of our work, it is the case that any block-diagonal correlations (as in Fig. \ref{fig:framework}) suffice as long as the $2\times 2$ ``un-normalized" correlations $C_{x,y,m}$ and $D_{x,y,m}$ imply the existence of reflections $Z_A, X_A$ on Alice's side and $Z_B, X_B$ on Bob's side such that \begin{align}
Z_A\ket{\psi}&=Z_B\ket{\psi} \label{f1}\\
X_A(\Id-Z_A)\ket{\psi}&=\tan{(\theta)}X_B(\Id+Z_A)\ket{\psi}  \label{f2}
\end{align}
for appropriate angles $\theta$.
For instance, in order to self-test bipartite maximally entangled states, we can invoke any correlation in the class given by Wang et al. \cite{Wang2016} where $A_0\ket{\psi}=B_0\ket{\psi}$ (in the notation of Ref \cite{Wang2016}, $\alpha_{00}=0$). These correlations satisfy equations \eqref{f1} and \eqref{f2} for $\tan\theta = 1$: thus, they can be used to self-test the maximally entangled pair of qudits, for any $d$, as is suggested by Yang and Navascu\'es \cite{Yang2013}.
For these correlations, notice, moreover, that for $x=0,y=0$, the correlation table is diagonal and hence, we can drop Bob's fourth measurement setting because a diagonal correlation can fulfil its role as both $C_{x,y,m}$ and $D_{x,y,m}$. Thus, one can self-test maximally entangled states of arbitrary dimension within a $[\{3,d\},\{3,d\}]$ Bell scenario.

We remark, that our analysis in the present work is limited to exact correlations: a natural follow-up to our work is to derive robustness bounds on our self-tests. While we believe that some robustness bounds can be derived, existing analytical tools produce notoriously unsatisfying bounds, and the numerical tools that give much better bounds can only be applied to few examples. In this situation, we'd rather wait for the progress in analytical tools of the kind found in \cite{Kaniewski2016}.

\section{Conclusion and outlook}
\label{conclusion}
In this work, we addressed the outstanding open question of whether all bipartite entangled quantum states can be self-tested. 
We presented a framework inspired by the work of Yang and Navascu\'es \cite{Yang2013}, and provided explicit correlations that self-test all bipartite entangled pure states, thus answering the question affirmatively. These are indeed all the bipartite states that one can hope to self-test, since any correlations achieved by mixed states can also be achieved by pure states of the same dimension \cite{Sikora2015}.
Our work provides new flexibility in choosing a bipartite quantum state and corresponding $d$-outcome ideal measurements for proofs of certification of randomness and of quantum computing; it also opens to the possibility of generating up to $\log(d)$ bits of private randomness with a seed consisting of two random trits per run, in a single Bell experiment setup. We leave this exploration for future work.

Other interesting directions/questions that remain open are the following:
\begin{itemize}
\item Is it possible to phrase the self-test of all bipartite entangled states in terms of maximal violation of a family of Bell inequalities?
\item Is there a family of non-local games whose optimal quantum values self-test all bipartite entangled states? We are currently aware of only one example of a non-local game self-testing a state that is not maximally entangled. It would be interesting to identify precisely what are the properties that non-local games must have in order to self-test states that are not maximally entangled. 

\end{itemize}

\section*{Acknowledgements} 
We thank Matthew McKague and Thomas Vidick for comments on earlier drafts, and acknowledge discussions with them as well as with Miguel Navascu\'es, Jalex Stark and Xingyao Wu.

This research is supported by the Singapore Ministry of Education Academic Research Fund Tier 3 (Grant No. MOE2012-T3-1-009); by the National Research Fund and the Ministry of Education, Singapore, under the Research Centres of Excellence programme. A.C. is supported by AFOSR YIP award number FA9550-16-1-0495.
\bibliographystyle{apalike}
\bibliography{references}

\begin{thebibliography}{}

\bibitem[Ac\'{\i}n et~al., 2007]{Acin2007}
Ac\'{\i}n, A., Brunner, N., Gisin, N., Massar, S., Pironio, S., and Scarani, V.
  (2007).
\newblock {Device-Independent Security of Quantum Cryptography against
  Collective Attacks}.
\newblock {\em Phys. Rev. Lett.}, 98:230501.

\bibitem[Ac\'{\i}n et~al., 2002]{Acin2002}
Ac\'{\i}n, A., Durt, T., Gisin, N., and Latorre, J.~I. (2002).
\newblock {Quantum nonlocality in two three-level systems}.
\newblock {\em Phys. Rev. A}, 65:052325.

\bibitem[Ac\'{\i}n et~al., 2012]{Acin2012}
Ac\'{\i}n, A., Massar, S., and Pironio, S. (2012).
\newblock {Randomness versus Nonlocality and Entanglement}.
\newblock {\em Phys. Rev. Lett.}, 108:100402.

\bibitem[Bamps and Pironio, 2015]{Bamps2015}
Bamps, C. and Pironio, S. (2015).
\newblock {Sum-of-squares decompositions for a family of
  Clauser-Horne-Shimony-Holt-like inequalities and their application to
  self-testing}.
\newblock {\em Phys. Rev. A}, 91:052111.

\bibitem[Bell, 1964]{Bell1964}
Bell, J.~S. (1964).
\newblock {On the einstein podolsky rosen paradox}.
\newblock {\em Physics (Long Island City, N.Y.)}, 1:195.

\bibitem[Chao et~al., 2016]{CRSV2016}
Chao, R., Reichardt, B.~W., Sutherland, C., and Vidick, T. (2016).
\newblock {Test for a large amount of entanglement, using few measurements}.
\newblock {\em arXiv preprint arXiv:1610.00771}.

\bibitem[Coladangelo, 2016]{Coladangelo2016}
Coladangelo, A.~W. (2016).
\newblock {Parallel self-testing of (tilted) EPR pairs via copies of (tilted)
  CHSH}.
\newblock {\em arXiv preprint arXiv:1609.03687}.

\bibitem[Coudron and Natarajan, 2016]{Coudron2016}
Coudron, M. and Natarajan, A. (2016).
\newblock {The Parallel-Repeated Magic Square Game is Rigid}.
\newblock {\em arXiv preprint arXiv:1609.06306}.

\bibitem[Coudron and Yuen, 2013]{Coudron2013}
Coudron, M. and Yuen, H. (2013).
\newblock {Infinite Randomness Expansion and Amplification with a Constant
  Number of Devices}.
\newblock {\em arXiv preprint arXiv:1310.6755}.

\bibitem[Kaniewski, 2016]{Kaniewski2016}
Kaniewski, J. (2016).
\newblock {Analytic and Nearly Optimal Self-Testing Bounds for the
  Clauser-Horne-Shimony-Holt and Mermin Inequalities}.
\newblock {\em Phys. Rev. Lett.}, 117:070402.

\bibitem[Kempe and Vidick, 2010]{KV10}
Kempe, J. and Vidick, T. (2010).
\newblock {Parallel Repetition of Entangled Games}.
\newblock {\em arXiv preprint arXiv:1012.4728}.

\bibitem[Mayers and Yao, 2004]{Mayers2004}
Mayers, D. and Yao, A. (2004).
\newblock {Self-testing quantum apparatus}.
\newblock {\em Quantum Inf. Comput.}, 4:273.

\bibitem[McKague, 2011]{McKague2011}
McKague, M. (2011).
\newblock {Self-testing graph states}.
\newblock In {\em Conference on Quantum Computation, Communication, and
  Cryptography}, pages 104--120. Springer.

\bibitem[McKague, 2016]{Mckague2016}
McKague, M. (2016).
\newblock {Self-testing in parallel}.
\newblock {\em New Journal of Physics}, 18(4):045013.

\bibitem[McKague et~al., 2012]{MYS12}
McKague, M., Yang, T.~H., and Scarani, V. (2012).
\newblock {Robust Self Testing of the Singlet}.
\newblock {\em J. Phys. A: Math. Theor.}, 45:455304.

\bibitem[Miller and Shi, 2016]{MillerShi2016}
Miller, C.~A. and Shi, Y. (2016).
\newblock {Robust protocols for securely expanding randomness and distributing
  keys using untrusted quantum devices}.
\newblock {\em Journal of the ACM}, 63:33.

\bibitem[Natarajan and Vidick, 2016]{Natarajan2016}
Natarajan, A. and Vidick, T. (2016).
\newblock {Robust self-testing of many-qubit states}.
\newblock {\em arXiv preprint arXiv:1610.03574}.

\bibitem[P{\'a}l et~al., 2014]{Pal2014}
P{\'a}l, K.~F., V{\'e}rtesi, T., and Navascu{\'e}s, M. (2014).
\newblock {Device-independent tomography of multipartite quantum states}.
\newblock {\em Phys. Rev. A}, 90(4):042340.

\bibitem[Pironio et~al., 2010]{Pironio2010}
Pironio, S., Ac{\'\i}n, A., Massar, S., de~la Giroday, A.~B., Matsukevich,
  D.~N., Maunz, P., Olmschenk, S., Hayes, D., Luo, L., Manning, T.~A., and
  Monroe, C. (2010).
\newblock {Random numbers certified by Bell's theorem}.
\newblock {\em Nature}, 464(7291):1021--1024.

\bibitem[Popescu and Rohrlich, 1992]{Popescu92}
Popescu, S. and Rohrlich, D. (1992).
\newblock {Which states violate Bell’s inequality maximally?}
\newblock {\em Phys. Lett. A}, 169:411--414.

\bibitem[Reichardt et~al., 2013]{Reichardt2013}
Reichardt, B.~W., Unger, F., and Vazirani, U. (2013).
\newblock {Classical command of quantum systems}.
\newblock {\em Nature}, 496(7446):456--460.

\bibitem[Salavrakos et~al., 2016]{Salavrakos2016}
Salavrakos, A., Augusiak, R., Tura, J., Wittek, P., Ac{\'\i}n, A., and Pironio,
  S. (2016).
\newblock {Bell inequalities for maximally entangled states}.
\newblock {\em arXiv preprint arXiv:1607.04578}.

\bibitem[Sikora et~al., 2016]{Sikora2015}
Sikora, J., Varvitsiotis, A., and Wei, Z. (2016).
\newblock {Minimum Dimension of a Hilbert Space Needed to Generate a Quantum
  Correlation}.
\newblock {\em Phys. Rev. Lett.}, 117:060401.

\bibitem[Summers and Werner, 1987]{SW87}
Summers, S.~J. and Werner, R. (1987).
\newblock {Bell's inequalities and quantum field theory. I. General setting.}
\newblock {\em J. Math. Phys.}, 28:2440.

\bibitem[Wang et~al., 2016]{Wang2016}
Wang, Y., Wu, X., and Scarani, V. (2016).
\newblock {All the self-testings of the singlet for two binary measurements}.
\newblock {\em New Journal of Physics}, 18(2):025021.

\bibitem[Wu et~al., 2016]{Wu2016}
Wu, X., Bancal, J.-D., McKague, M., and Scarani, V. (2016).
\newblock {Device-independent parallel self-testing of two singlets}.
\newblock {\em Phys. Rev. A}, 93(6):062121.

\bibitem[Wu et~al., 2014]{Wu2014}
Wu, X., Cai, Y., Yang, T.~H., Le, H.~N., Bancal, J.-D., and Scarani, V. (2014).
\newblock {Robust self-testing of the three-qubit W state}.
\newblock {\em Phys. Rev. A}, 90(4):042339.

\bibitem[Yang and Navascu\'es, 2013]{Yang2013}
Yang, T.~H. and Navascu\'es, M. (2013).
\newblock {Robust self-testing of unknown quantum systems into any entangled
  two-qubit states}.
\newblock {\em Phys. Rev. A}, 87:050102.

\bibitem[Yang et~al., 2014]{Yang2014}
Yang, T.~H., V\'ertesi, T., Bancal, J.-D., Scarani, V., and Navascu\'es, M.
  (2014).
\newblock {Robust and Versatile Black-Box Certification of Quantum Devices}.
\newblock {\em Phys. Rev. Lett.}, 113:040401.

\end{thebibliography}

\newpage

\begin{appendix}
\section{Proof of Lemma \ref{YNcriterion}} 
\label{YNproof}
In this section, we provide a proof of Lemma \ref{YNcriterion}. We explicitly construct a local isometry $\Phi$ such that $\Phi(\ket{\psi})=\ket{\text{extra}}\otimes\ket{\psi_{\text{target}}}$, where the ideal target state is $\ket{\psi_{\text{target}}}=\sum_{i=0}^{d-1} c_i \ket{ii}$, and $\ket{\text{extra}}$ is some auxiliary state. 

\begin{proof}
Recall that $\{P_A^{(k)}\}$ is a complete orthogonal set of orthogonal projections by hypothesis. Then, notice that for $i \neq j$ we have, using condition \eqref{c2}, $P_B^{(i)}P_B^{(j)} \ps = P_B^{(i)}P_A^{(j)} \ps =P_A^{(j)}P_A^{(i)} \ps = 0$, i.e the $P_B^{(k)}$ are ``orthogonal when acting on $\ps$''. Then, we can invoke a slight variation of the \textit{orthogonalization lemma} (Lemma 21 from Kempe and Vidick \cite{KV10}) to obtain projections on Bob's side that are exactly orthogonal, and have the same action on $\ps$.
\begin{lem}
\label{projections}
Let $\rho$ be positive semi-definite, living on a finite-dimensional Hilbert space. Let $P_1,..,P_k$ be projections such that 
\begin{equation}
\sum_{i\neq j} \tr(P_iP_jP_i\rho) \leq \epsilon
\end{equation}
for some $0<\epsilon \leq Tr(\rho)$. Then there exist orthogonal projections $Q_1,..,Q_k$ such that
\begin{equation}
\sum\limits_{i=1}^{k} \tr\big((P_i-Q_i)^2\rho\big) =O\big(\epsilon^{\frac12}\big)\tr(\rho)^{\frac12}
\end{equation}

\end{lem}
This gives us a new set of orthogonal projections $\{\tilde{P}_B^{(k)}\}$ such that $\tilde{P}_B^{(k)} \ps =P_B^{(k)} \ps \,\,\forall k$. 

Now, define $Z_{A} := \sum_{k=0}^{d-1} \omega^k P_A^{(k)}$ and $Z_{B} := \sum_{k=0}^{d-1} \omega^k \tilde{P}_{B}^{(k)} + \mathds{1} - \sum_{k=0}^{d-1} \tilde{P}_{A/B}^{(k)}$.
In particular, $Z_A$ and $Z_B$ are unitary. Notice, moreover, that $\big(\mathds{1}-\sum_{k=0}^{d-1} \tilde{P}_{A/B}^{(k)}\big) \ps = 0$, again using condition \eqref{c2}.

Define the local isometry
\begin{equation}
\Phi := (R_{AA'}\otimes R_{BB'})(\bar{F}_{A'}\otimes\bar{F}_{B'})(S_{AA'}\otimes S_{BB'})(F_{A'}\otimes F_{B'})
\end{equation}
where $F$ is the quantum Fourier transform, $\bar{F}$ is the inverse quantum Fourier transform, $R_{AA'}$ is defined so that $\ket{\phi}_{A} \ket{k}_{A'} \mapsto X^{(k)}_{A}\ket{\phi}_{A} \ket{k}_{A'}\,\,\, \forall \ket{\phi}$, and similarly for $R_{BB'}$, and $S_{AA'}$ is defined so that $\ket{\phi}_{A} \ket{k}_{A'} \mapsto Z^{k}_{A}\ket{\phi}_{A}\ket{k}_{A'} \,\,\, \forall \ket{\phi}$, and similarly for $S_{BB'}$. We compute the action of $\Phi$ on $\ket{\psi}_{AB}\ket{0}_{A'}\ket{0}_{B'}$. For ease of notation with drop the tildes from the $\tilde{P}_B^{(k)}$, while still referring to the new orthogonal projections. 
\begingroup
\allowdisplaybreaks
\begin{align}
\ket{\psi}_{AB}\ket{0}_{A'}\ket{0}_{B'} \stackrel{F_{A'}\otimes F_{B'}}{\longrightarrow}& \frac{1}{d}\sum_{k,k'}\ket{\psi}_{AB}\ket{k}_{A'}\ket{k'}_{B'} \\
\stackrel{S_{AA'}\otimes S_{BB'}}{\longrightarrow}& \frac{1}{d}\sum_{k,k'}\left(\sum_{j}\omega^{j}P^{(j)}_{A}\right)^{k}\left(\sum_{j'}\omega^{j'}P^{(j')}_{B} + \mathds{1} - \sum_k P_B^{(j')}\right)^{k'}\ket{\psi}_{AB}\ket{k}_{A'}\ket{k'}_{B'}\\
=& \frac{1}{d}\sum_{k,k',j,j'}\omega^{jk}\omega^{j'k'}P^{(j)}_{A}P^{(j')}_{B}\ket{\psi}_{AB}\ket{k}_{A'}\ket{k'}_{B'}\\
=& \frac{1}{d}\sum_{k,k',j,j'}\omega^{jk}\omega^{j'k'}P^{(j)}_{A}P^{(j')}_{A}\ket{\psi}_{AB}\ket{k}_{A'}\ket{k'}_{B'}\\
=& \frac{1}{d}\sum_{k,k',j}\omega^{j(k+k')}P^{(j)}_{A}\ket{\psi}_{AB}\ket{k}_{A'}\ket{k'}_{B'}\\
\stackrel{\bar{F}_{A'}\otimes \bar{F}_{B'}}{\longrightarrow}&\frac{1}{d^2}\sum_{k,k',j,l,l'}\omega^{j(k+k')}\omega^{-lk}\omega^{-l'k'}P^{(j)}_{A}\ket{\psi}_{AB}\ket{l}_{A'}\ket{l'}_{B'}\\
=&\frac{1}{d^2}\sum_{k,k',j,l,l'}\omega^{k(j-l)}\omega^{k'(j-l')}P^{(j)}_{A}\ket{\psi}_{AB}\ket{l}_{A'}\ket{l'}_{B'}\\
=& \sum_{j}P^{(j)}_{A}\ket{\psi}_{AB}\ket{j}_{A'}\ket{j}_{B'}\label{A12}\\
\stackrel{R_{AA'}\otimes R_{BB'}}{\longrightarrow}& \sum_{j}X^{(j)}_{A}X^{(j)}_{B}P^{(j)}_{A}\ket{\psi}_{AB}\ket{j}_{A'}\ket{j}_{B'}\\
=& \sum_{j}\frac{c_j}{c_0} P^{(0)}_{A}\ket{\psi}_{AB}\ket{j}_{A'}\ket{j}_{B'}\\
=& \frac{1}{c_0}P^{(0)}_{A}\ket{\psi}_{AB} \otimes \sum_{j} c_j \ket{j}_{A'}\ket{j}_{B'}\\
=& \ket{\text{extra}} \otimes \ket{\psi_{\text{target}}}
\end{align}
\end{proof}
\endgroup
It is an easy check to see that the whole proof above can be repeated by starting from a mixed joint state, yielding a corresponding version of the Lemma that holds for a general mixed state.

\section{Self-testing the measurements}
\label{self-testing measurements}
Not much work is required to extend self-testing to the measurement operators, using the same local isometry $\Phi$, defined via the projections $P_{A/B}^{(k)}$ and the unitary operators $Z_{A/B}$ and $X^{(k)}_{A/B}$, as defined in the main text. 

Consider $\hat{A}_{x,m}=\Pi^{A_x}_{2m}-\Pi^{A_x}_{2m+1}$ and $\hat{B}_{y,m}=\Pi^{B_y}_{2m}-\Pi^{B_y}_{2m+1}$. Let $A_{x,m}$,$B_{y,m}$ be the two-qubit ideal measurements achieving maximal violation of tilted CHSH on the (2m,2m+1) subspace, i.e. $A_{0,m} = [\sigma_Z]_m$, $A_{1,m} = [\sigma_X]_m$, $B_{0,m} = [\cos(\mu_m)\sigma_Z + \sin(\mu_m)\sigma_X]_m$, $B_{1,m} = [\cos(\mu_m)\sigma_Z - \sin(\mu_m)\sigma_X]_m$, with the notation from subsection \ref{ideal measurements}. We claim, first, that $\Phi(\hat{A}_{x,m}\ket{\psi}) =  \ket{\text{extra}} \otimes A_{x,m} \ket{\psi_{\text{target}}} $ and $\Phi(\hat{B}_{y,m}\ket{\psi}) =  \ket{\text{extra}} \otimes B_{y,m} \ket{\psi_{\text{target}}} $. \\
Following closely the proof in Appendix \ref{YNproof} up to Equation \eqref{A12}, we have 
\begin{align}
\Phi(\hat{A}_{x,m}\ket{\psi}) &= R_{AA'}\otimes R_{BB'} \sum_{j}P^{(j)}_{B}\hat{A}_{x,m}\ket{\psi}_{AB}\ket{j}_{A'}\ket{j}_{B'} \nonumber \\
&= R_{AA'}\otimes R_{BB'} \Big(P^{(2m)}_{B}\hat{A}_{x,m}\ket{\psi}_{AB}\ket{2m}_{A'}\ket{2m}_{B'}+ P^{(2m+1)}_{B}\hat{A}_{x,m}\ket{\psi}_{AB}\ket{2m+1}_{A'}\ket{2m+1}_{B'} \Big) \nonumber \\
&=X_{B}^{(2m)}X_{A}^{(2m)} P^{(2m)}_{B}\hat{A}_{x,m}\ket{\psi}_{AB}\ket{2m}_{A'}\ket{2m}_{B'}+X_{B}^{(2m+1)}X_{A}^{(2m+1)}P^{(2m+1)}_{B}\hat{A}_{x,m}\ket{\psi}_{AB}\ket{2m+1}_{A'}\ket{2m+1}_{B'} \nonumber \\
&= X_{B}^{(2m)}X_{A}^{(2m)} \Big(P^{(2m)}_{B}\hat{A}_{x,m}\ket{\psi}_{AB}\ket{2m}_{A'}\ket{2m}_{B'}  +X_{A,m} X_{B,m}P^{(2m+1)}_{B}\hat{A}_{x,m}\ket{\psi}_{AB}\ket{2m+1}_{A'}\ket{2m+1}_{B'} \Big) \nonumber\\
&=X_{B}^{(2m)}X_{A}^{(2m)} \frac{1}{c_{2m}}P^{(2m)}_{B}\ket{\psi}_{AB}\otimes A_{x,m}\big(c_{2m}\ket{2m}_{A'}\ket{2m}_{B'}  +c_{2m+1}\ket{2m+1}_{A'}\ket{2m+1}_{B'}\big) \nonumber\\
&=  \frac{1}{c_0}P^{(0)}_{A}\ket{\psi}_{AB}  \otimes A_{x,m} \ket{\psi_{\text{target}}}  =  \ket{\text{extra}} \otimes A_{x,m} \ket{\psi_{\text{target}}} 
\end{align}
where the second-to-last line follows from the definitions of $X_{A,m}$ and $X_{B,m}$ in the main text, and from a proof following closely that in \cite{Bamps2015}, that maximal violation of the tilted CHSH inequality self-tests the ideal single-qubit measurements. One obtains analogous statements involving $\hat{A}'_{0/1,m}=\Pi^{A_{0/2}}_{2m+1}-\Pi^{A_{0/2}}_{2m+2}$ and $\hat{B}'_{0/1,m}=\Pi^{B_{2/3}}_{2m+1}-\Pi^{B_{2/3}}_{2m+2}$.\\
From the above, we deduce that the measurements of Alice and Bob on $\ps$ are equivalent under $\Phi$, to the ideal measurements described in subsection \ref{ideal measurements} on $\ket{\psi_{\text{target}}} $. 

\end{appendix}

\end{document}